%% file: ms.tex
\documentclass{ws-fnl}
\usepackage{cite}
\usepackage{url}

\newcommand{\pref}{\protect\ref}
\newcommand{\solrad}{\ifmmode{R}_{\rm S}\else${R}_{\rm S}$\fi}
\newcommand{\solmas}{\ifmmode{M}_{\rm S}\else${M}_{\rm S}$\fi}

\newcommand{\flxu}{\ifmmode{\rm W~m^{-2}}\else W~m$^{-2}$\fi}
\newcommand{\intu}{\ifmmode{\rm W~m^{-2}~sr^{-1}~hz^{-1}}\else 
                   W~m$^{-2}$~sr$^{-1}$~hz^{-1}\fi}
\newcommand{\tintu}{\ifmmode{\rm W~m^{-2}sr^{-1}}\else W~m$^{-2}$~sr$^{-1}$\fi}

\newcommand{\wave}{\ifmmode{\lambda} \else$\lambda$\fi}

\def\lesssim{\mathrel{\hbox{\rlap{\hbox{\lower4pt\hbox{$\sim$}}}\hbox{$<$}}}}

\def\ion#1#2{#1$\;${\rm%
      \expandafter\uppercase\expandafter{\romannumeral #2}\ignorespaces}\relax}

\renewcommand{\vec}[1]{{\bf #1}}
\newcommand{\matrx}[1]{
    \ifmmode{\underline{\underline {\bf #1}}}
    \else ${\underline{\underline {\bf #1}}}$
    \fi}
\newcommand{\totd}[1]{
    \ifmmode{ {{\partial{#1}}\over {\partial t}} + \vec{v}\cdot\nabla #1}
    \else${ {{\partial{#1}}\over {\partial t}} + \vec{v}\cdot\nabla #1}$
    \fi}

\newcommand{\grad}{{\bf grad\,}}
\newcommand{\curl}{{\bf curl\,}}

\newcommand{\cond}{\ensuremath \sigma}

\newcommand{\B}{\vec{B}}
\newcommand{\jj}{\vec{j}}

\newcommand{\dd}{\partial}

\newcommand{\dt}{\partial t}

\newcommand{\avr}[1]{\overline{#1}}
\newcommand{\pr}[1]{{#1}^{\prime}}

\def\mathstacksym#1#2#3#4#5{\def#1{\mathrel{\hbox to 0pt{\lower {#3}\hss} \raise #4\hbox{#2}}}}

\mathstacksym\lta{$<$}{$\sim$}{1.5pt}{3.5pt} 
\mathstacksym\gta{$>$}{$\sim$}{1.5pt}{3.5pt} 
\mathstacksym\lrarrow{$\leftarrow$}{$\rightarrow$}{2pt}{1pt} 
\mathstacksym\lessgreat{$>$}{$<$}{3pt}{3pt} 

\begin{document}

\title{The mercurial Sun at the heart of our solar system}

\markboth{The Sun}{P.G. Judge}

\author{Philip G. Judge}

\newcommand{\hao}{
High Altitude Observatory,\\
National Center for Atmospheric Research,\\
Boulder CO 80307-3000,
 USA}
 
\address{\hao,\\ judge@ucar.edu}

\maketitle

\begin{abstract}
  As the powerhouse of our solar system, the Sun's electromagnetic planetary influences
appear contradictory. On the one hand, the Sun for aeons emitted radiation which was  ``just
right" for life to evolve in our terrestrial Goldilocks zone, even for
such complex organisms as ourselves.  On the other, in the dawn
of Earth's existence the Sun was far dimmer than today, and yet evidence
for early liquid water is written into geology.  Now in middle age,
the Sun should be a benign object of little
interest to society or even astronomers.  However, for  physical reasons yet to be fully
understood, it contains a magnetic machine with a slightly arrhythmic 11 year magnetic heartbeat.   Although these variations require  merely  
 0.1\%{} of the solar luminosity, 
 this power floods the solar system with rapidly changing fluxes of photons and particles at energies far above the 0.5eV thermal energy characteristic of the photosphere. Ejected solar plasma carries magnetic fields into  space with consequences for planets, the Earth being vulnerable to geomagnetic storms. 
 This chapter  discusses some physical reasons why the
Sun suffers from such ailments, and examine consequences through time
across the solar system. A Leitmotiv of the discussion is that any rotating
and convecting star must inevitably generate magnetic ``activity'' for which the Sun
represents the example \textit{par excellence}.
\end{abstract}

\section*{Introduction}

The present paper attempts to
introduce astronomers concerned with exo-planetary studies to electromagnetic 
solar influences on its planetary system.
The approach follows that of a recent book 
\cite{Judge2020}, presenting a viewpoint of a physicist and astronomer,
not of a solar 
specialist.  
\begin{figure}[ht]
\begin{center}
\includegraphics[width=0.9\linewidth]{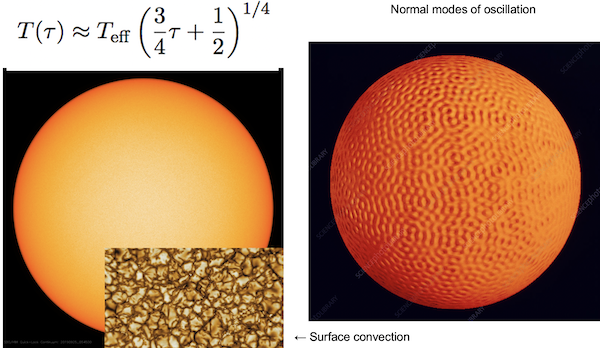}
\end{center}
\caption{A fictional 
star without magnetic fields. The equation shows the run of
temperature with grey optical depth $\tau$ for a radiative equilibrium atmosphere. 
The left panel shows limb darkening and surface convection, with
roughly 4 million
granules on the solar surface at any moment.
The right panel shows Doppler shifts associated with the incoherent superposition of normal modes of oscillation. The solar rotation is seen as an East-West gradient, spanning  $\pm$ 2 km~s$^{-1}$.
}
\label{fig:fiction}
\end{figure}
  Table 1 lists gross solar properties, which could be used together with elemental abundances and data from elementary  physics, to construct 
a theoretical star of the kind shown in Figure~\ref{fig:fiction}.  But this fictional star merely reflects a star without magnetic fields.  Magnetism arises from the differential motions of ions and electrons within plasmas like the Sun, and because there are no magnetic charges (monopoles), magnetic fields are not shorted out, and can pervade plasmas. 
\begin{figure}[ht]
\begin{center}
\includegraphics[width=\linewidth]{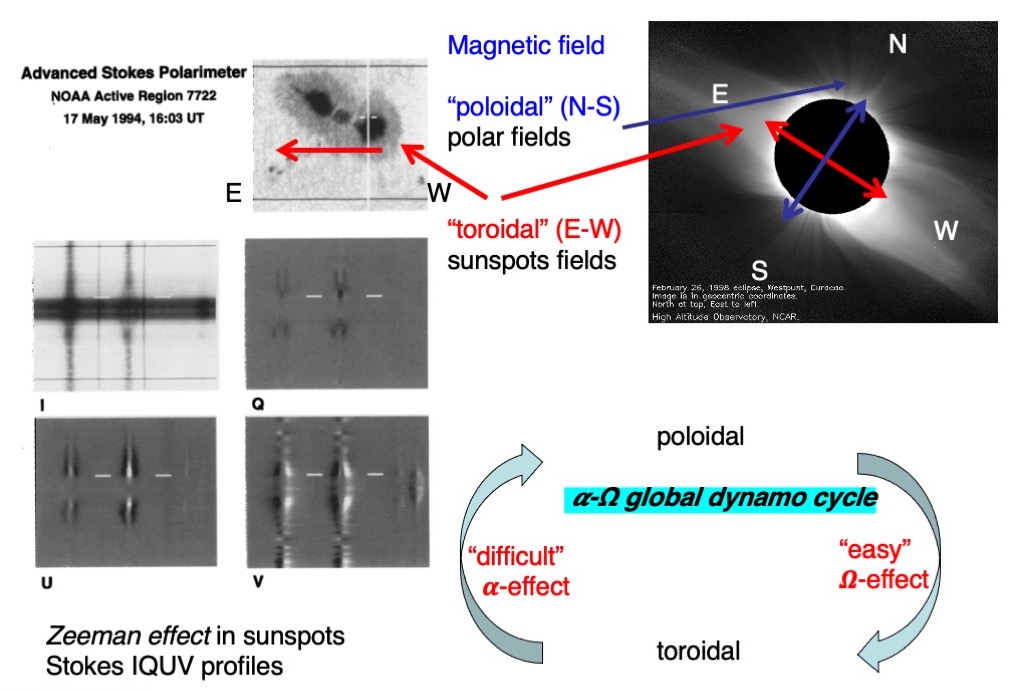}
\end{center}
\caption{The images reveal observational signatures of solar magnetic fields.
Outside of sunspots magnetism is best detected
through the polarization of light by the Zeeman effect.   
Sunspots themselves are aligned 
mostly E-W and are believed to 
originate from the emergence of tori of magnetic field 
generated by differential rotation of
fluid in the solar convection zone
($\Omega$-effect).   Polar magnetic fields are traced by coronal plasma 
along N-S oriented rays in the coronal image. The schematic ``dynamo cycle'' of magnetic field is illustrated demonstrating the continuous process of generating poloidal fields (seen at the pole) from toroidal fields (sunspots), and vice versa.
}
\label{fig:cycle}
\end{figure}
In contrast, large-scale electric fields cannot
be supported in plasmas.  The interactions between solar plasma and magnetism constitutes the main focus of interest in modern solar physics.   The clearest signature of magnetism in stars and the Sun is in spots (Figure~\ref{fig:cycle}). Not only are these much darker than the granulated surface, but 
by using polarized light the magnetic flux can be measured through the Zeeman effect (a technique started by Hale and colleagues  \cite{Hale1908,Hale+others1919}).

\begin{figure}[ht]
\begin{center}
\includegraphics[width=\linewidth]{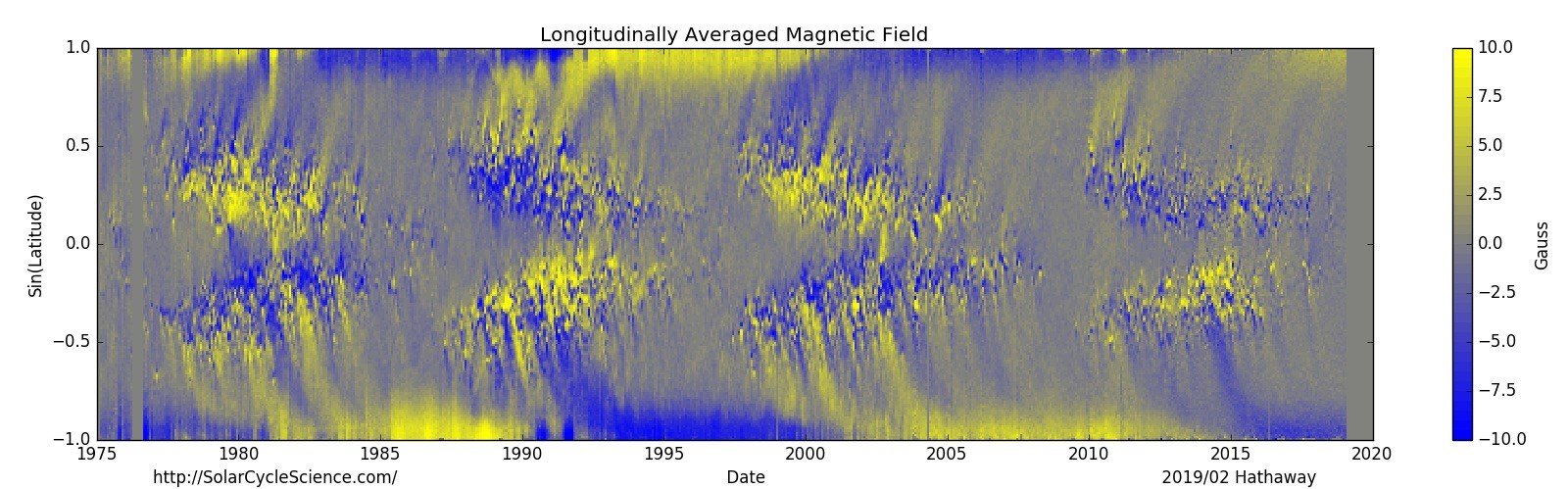}
\end{center}
\caption{The signed surface  magnetic flux densities (units Mx~cm$^{-2}$) are shown as a function of time and latitude,  derived from Stokes $V$ profiles of spectral lines (lower right spectral image in Figure~\ref{fig:cycle}).  Poloidal components are seen as the yellow and blue patches
close to the two poles, and the (mostly) toriodal fields are seen in the ``butterfly wings'' which are mostly oriented E-W (Figure~\ref{fig:cycle}).
Between these, surface fields   
propagate towards the poles, mostly of opposite polarity to the polar fields, appearing to reverse the polarity of polar fields every 11 years. The butterfly wing pattern was originally discovered by Annie and Edward Maunder in 1904. }
\label{fig:hathaway}
\end{figure}

The present discussion encompasses the Sun's behavior as it affects the solar system, from
zero-age main sequence to the present day and beyond. On timescales short compared with thermonuclear processing, this 
magnetic ``activity'' is responsible for electromagnetic disturbances perturbing the planets.  
Over 5 decades, solar and stellar studies reveal that  magnetically-induced solar variations 
associated with sunspots are not exceptional among the stars. It is, however, somewhat \textit{mercurial}\footnote{A character described as  changeable; volatile;  lively; sprightly;  fickle; flighty; erratic. From Greek mythology.}:  
on the one hand the well-known sunspot cycle  (an aspect of which is shown in Figure~\ref{fig:hathaway}) is  the most regular 11-year cyclic variation of all stars
\cite{Egeland2017}. On the other, these cycles are punctuated by irregular epochs of weakened magnetic influences on timescales of centuries. Dynamic,  unpredictable flares often associated with coronal mass ejections (CMEs) vary over
timescales of minutes
\cite{Eddy2009}.
CMEs are  magnetized bubbles of energetic plasma 
into interplanetary
space unleashed by the Sun,
frequently with consequences for a society dependent on a stable global electromagnetic environment
\cite{Eddy2009}.  
The precise mechanisms underlying  this behavior remain  beyond current understanding.   However, some of the ingredients that are known 
are discussed in an Appendix, acknowledging that 
non-linear and non-local  physical effects 
still present us with
significant fundamental challenges.  Thus our understanding
is still largely driven by observations and not from consideration of first-principles. This argument applies both to the regeneration of global solar magnetism discussed in the appendix, as well as 
effects such as coronal heating and dynamics (e.g., 
\cite{Judge2021pw}.)
\begin{center}
\textbf{Table 1.   Some basic properties of the Sun}
  \begin{tabular}{lll}
   \hline
Age & 4.54 Gyr \\
  Mass $M_\odot$ & $2\times10^{30}$ kg\\
  Radius $R_\odot$ & 700,000 km\\
  Distance & 150,000,000 km $\equiv$ 1 AU\\
  Luminosity $L_\odot$ & $4\times10^{26}$ W\\
  Irradiance at Earth  & 1.365 kW m$^{-2}$\\
  Average rotation period$^\ast$ & 27 days \\
  Tilt of rotation axis to ecliptic& 7$^\circ$\\
  Stellar spectral type & G2 V\\
  $B-V$ color & 0.65\\
  Absolute magnitude $M_V$ & 4.83\\
  \hline
\end{tabular}
\end{center}
{\footnotesize$^\ast$The surface rotates differentially, from 25 days at the equator to over 30 days near the poles.}

\section*{A non-magnetic Sun}
A fictional Sun-like star without magnetic fields 
(Figure~\ref{fig:fiction}) 
has benign influences on its planetary system:
\begin{itemize}
    \item The star  brightens on the main sequence by 
    just 13\% per Gyr, with a ZAMS effective temperature $T_\mathrm{eff}$ of 5660 K, increasing by $\approx 25$ K per Gyr. 
    \item The planets are irradiated by a near black-body spectrum with temperature near $T_\mathrm{eff}$.
    \item These
    irradiances would be weakly modulated by 
    small amplitude random variations of surface convection (granules) on periods between 2 and 10 minutes, and the linear  oscillations of normal modes of oscillation of the elastic sphere, peaked near periods of 5 minutes \cite{Ulrich1970}. 
\end{itemize}
This fictional star would provide a stable input to interplanetary space over aeons, conducive to the slow evolution of life, necessary perhaps  to develop complex multi-cellular life like ourselves \cite{Catling2013}.
It would  possess no reservoir to store free energy
in its atmosphere outside of those in fluid wave modes, and emit almost no UV or X- radiation. 
\begin{figure}[ht]
\begin{center}
\includegraphics[width=0.8\linewidth]{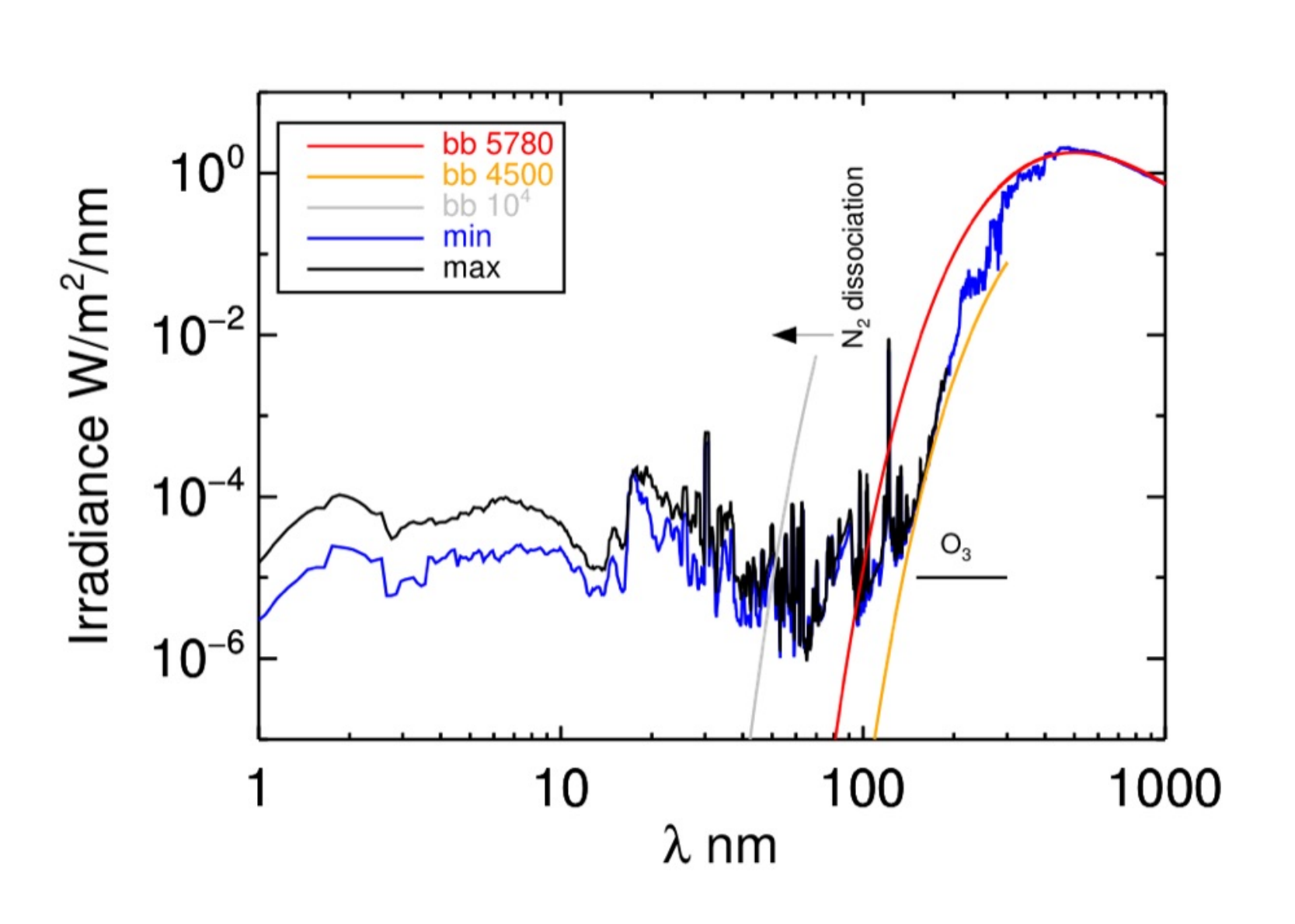}
\end{center}
\caption{Five flux spectra 
relevant to Sun-like stars 
are shown. Smooth lines show black body spectra. 
The blue line 
is a reference 
spectrum measured 
close to minimal solar
magnetic activity
\cite{lisird},
the black a representative spectrum close to  sunspot maximum.
 The 
arrow indicates wavelengths at which the N$_2$ molecule, dominant in the thermosphere, is photo-dissociated, indicating that a non-magnetic Sun 
would possess no permanent ionosphere. However,  stratospheric ozone (O$_3$) would be formed in either case.
}
\label{fig:spectrum}
\end{figure}
In stark contrast,
Figure~\ref{fig:spectrum} also shows solar spectra
between the maxima and minima of the 11 year cycle of sunspots, as black and blue lines respectively.  No solar magnetism means no ionosphere.  
 
 \section*{Magnetic Sun}
The reasons why the Sun must behave
according to 
Figures~\ref{fig:cycle}-\ref{fig:spectrum} are
 by no means obvious, either from a 
 theoretical (see the Appendix) or empirical point of view.  For example, there exist rapidly-rotating Sun-like stars without spot cycles, with more energetic  magnetic fields  and lacking the level of symmetry exhibited in time and space by the Sun in Figure~\ref{fig:hathaway}.  Sun-like stars can show 
 spots on rotational time scales (weeks)
 but with no sunspot
 cycles (e.g., \cite{Egeland2017}).

\begin{figure}[ht]
\begin{center}
\includegraphics[width=0.9\linewidth]{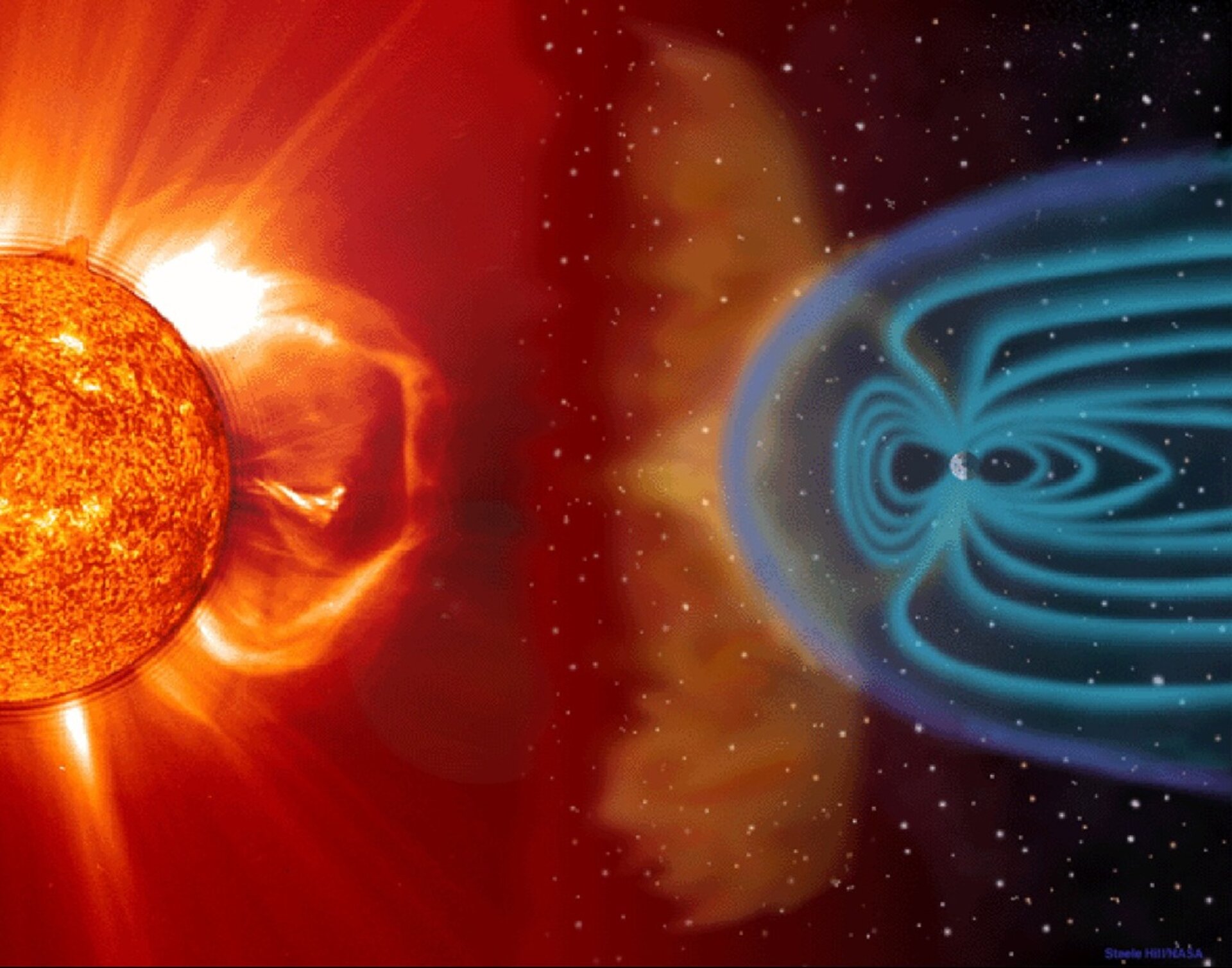}
\end{center}
\caption{Two images of the real Sun observed by instruments on the SoHO spacecraft 
are shown (left) along with a sketch 
(not to scale) of the Earth and its magnetosphere (right).   A non-magnetic star 
would show neither
the solar disk nor corona, 
 could not produce
energetic particles 
(white dots) or a mass ejection, and the Earth's magnetosphere would be unperturbed.  
Figure produced by ESA. 
}
\label{fig:sketch}
\end{figure}

The solar magnetic fields   
can usefully be projected on to poloidal and 
toroidal components (Figure~\ref{fig:cycle}, see also the Appendix), which are convenient also for modeling solar magnetic evolution (e.g., \cite{1999ApJ...518..508D}.)  
In the context of models described in the Appendix, the solar cycle involves 
the repeated interplay between the 
toroidal and poloidal fields on a time scale of 11 years. 
Some 
effects of these variable surface magnetic fields are highlighted in 
Figure~\ref{fig:sketch}, showing an image of
the solar disk seen in the resonance line
of He$^+$ (orange) at 30.4 nm. It is seen as a prominent peak along with 
the H L$\alpha$ line 
at 121.6 nm in the flux (irradiance) spectra  shown in  Figure~\ref{fig:spectrum}.  These lines arise from plasma close to 100,000 and 20,000 K respectively.  It also shows the solar corona close to 1 million K observed in
broad-band light (darker orange),  both images from instruments on the SoHO
spacecraft.  The dark orange ``ear-like''  structure is
a coronal mass ejection (CME) extending over a solar radius above the surface.  CMEs are
enormous, hot magnetic plasmoids arising from and separated from the Sun by magnetic reconnection \cite{Eddy2009,Priest+Forbes2007}.
The right half of the figure shows a sketch (not to scale) of the CME later as it has propagated through  interplanetary space (diffuse tan), energetic particles 
(white dots), the Earth's magnetosphere (blue), and the CME as it later impacts the magnetopause region of Earth's atmosphere.   The cartoon shows 
a quiescent phase before the blue (magnetospheric) and tan (CME) 
magnetic fields interact. More consequential are those phases where 
the CME plasma
gains entry into the 
Earth's magnetosphere through magnetic reconnection  at the interface region (``magnetopause'') and in the stretched out tail of the magnetosphere, causing    potentially damaging geomagnetic storms.

These phenomena all arise 
from the storage and release  of magnetic free energy in the corona, caused by 
the emergence and perturbation of magnetic fields generated by  dynamo action in the  solar interior \cite{Brun+Browning2017}. 
The measured irradiation of the solar system varies 
on all time scales from 
seconds (flares) to decades (the waxing and waning of sunspots).  
Representative variations over decades are shown in Figure~\ref{fig:spectrum}, showing typical solar spectra in 
 blue and black lines, 
representative of a minimum
and maximum in the number of spots on the Sun. These changes 
occur on a timescale of just 5 years!
The figure also shows 
that little of the entire 
solar luminosity is carried shortward of 200 nm, but that 
the variations 
in the irradiance 
increase systematically 
with decreasing wavelength.  
Soft X-rays of 1 keV energy (wavelengths near 1 nm) vary over decades almost by a factor of 10 as sunspots come and go. By comparison, the \textit{total}  (wavelength-integrated irradiance as measured at Earth) varies only by
about 0.1\%.  However, the passage of a single 
large sunspot group over several days can change the total irradiance for days by 1\%. Numbers of short-lived but highly energetic 
solar flares and CMEs are statistically correlated with sunspot numbers \cite{Ramesh2010ApJ}, during which the EUV and X-ray irradiances can increase by orders of magnitude 
over minutes and hours. Later we will see that the Sun can produce 
far more energetic flares than have yet been recorded.

In short, the differences between the fictional star and the Sun indicate that 
\begin{quote}
    \textit{the Sun is a  
machine which generates magnetic fields, with a strong 22 year periodicity.  The evolving magnetism emits 
high energy radiation
and plasma particles into interplanetary space, 
which fluctuate on 
timescales down to minutes.} 
\end{quote}
The Appendix summarizes 
what this remarkable solar behaviour implies about underlying physical causes.

\section*{The Sun, stars, and life on Earth}

We inhabit a planet around an ordinary, middle-aged star in the outer parts of an ordinary spiral galaxy. Even before the first detections of extrasolar planets, astronomical evidence (for example, in the distribution of elemental abundances), together with the 
remarkable success  of the theory of stellar evolution
(e.g., \cite{Burbidge+others1957}), suggests that 
the solar system  
is an ordinary and  natural product of 
stellar and galactic evolution. It arose from the debris of earlier generations of ordinary massive stars which lived their whole lives in the Galaxy.  In terms of universal
life, and assuming (as usual) that the laws of nature are universal, this mundane situation  raises  profound philosophical questions: is life nevertheless so rare that our solar system is the only place in the universe with life? Or is life teeming across the universe \cite{Catling2013}, in which case, where is everyone (a question attributed to Enrico Fermi in 1950)?

After Earth's formation, bombardments, tectonic 
activity and the acquisition of
water, the Earth 
entered a phase of relative stability  suitable for the early creation of life.  Threats to life over the following aeons include 
variations in global climate, asteroid and meteoroid collisions, and, perhaps, high energy events from the Sun.    
The remains of this discussion asks, 
\textit{how did the Sun's magnetic machinery evolve, and could there have been important consequences  for evolution of complex life on Earth?}  After all, high energy 
electromagnetic disturbances were
essential components leading to  ``primordial soup'' of amino acids  from inorganic ingredients in the 
famous Miller-Urey experiments in the 1950s. 

In this ``big 
picture'' view, I am struck by a poetic 
parallel which can be drawn between 
planetary and stellar
magnetism.
\begin{figure}[ht]
\begin{center}
\includegraphics[width=0.8\linewidth]{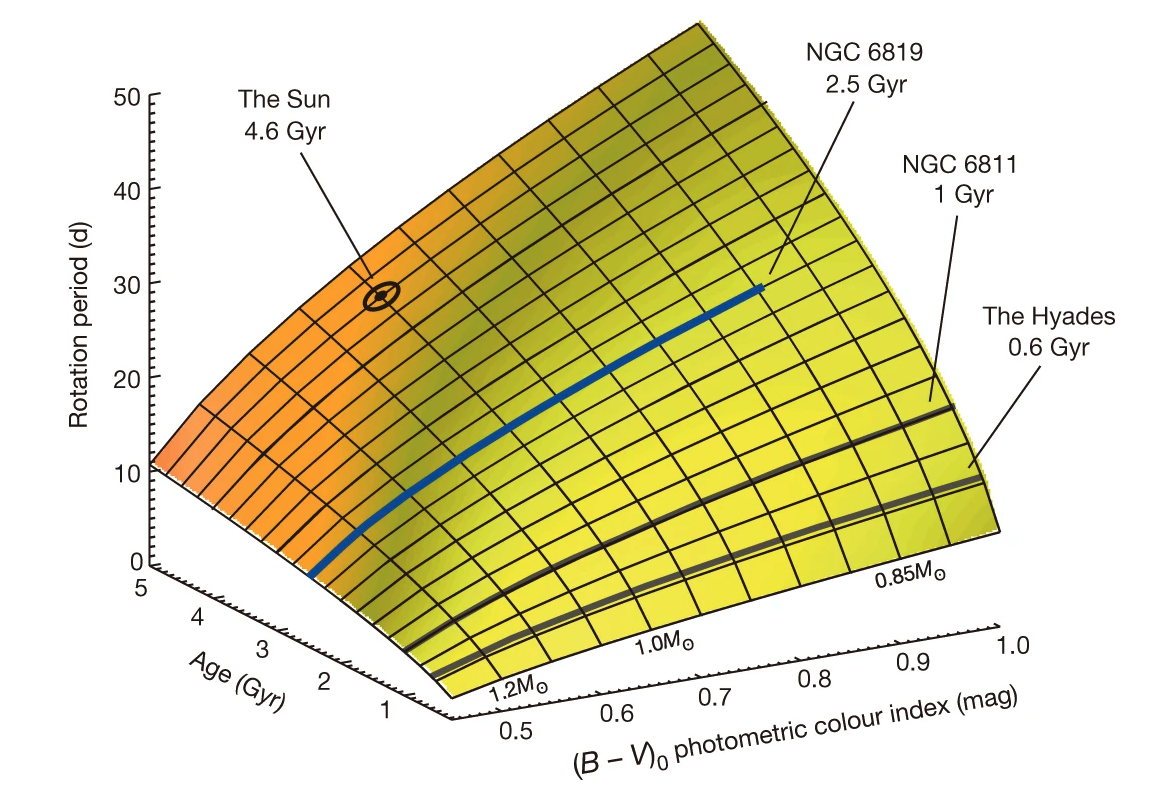}
\end{center}
\caption{A modern summary of 
relations between stellar rotation and age.   For each color index, a single curve along  the surface gives the empirical 
rotation - age relationship.  Measurements of rotation period can thus
be used to estimate stellar ages, a method termed ``gyro-chronology''.  Figure taken from \cite{Meibom+others2015}.
}
\label{fig:gyro}
\end{figure}
After the formation of the first stars, young massive stars converted
primordial H, He and a trace of 
Li into the array of elements 
generated both by fusion and neutron capture processes.  Thanks in part to the decay of radioactive elements like $^{232}$U in the  interior, the Earth's core is heated, and in part, molten and subject to convection. Coupled with Earth's rotation, Walter Els\"asser \cite{Elsasser1939}
first suggested that 
the convecting fluid then generates its own magnetic field.  
serving as a protective shield 
against hostile energetic 
charged particles of
cosmic and solar origins.   Thus, while the Sun's magnetised fluid generates threats, the Earth naturally   
generates a natural defense, through the same kind of
mechanism (a dynamo).
The story of stellar rotation and associated magnetic evolution on the main sequence 
is also, perhaps
a poem of epic 
proportions:
\begin{itemize}
\item Rotation and convection are 
natural consequences of 
the formation and  evolution of stars,
\item these are the 
essential ingredients to generate global and variable global 
magnetic fields (Appendix), then  
\item an increase in temperature is an unavoidable consequence 
of the motions of emerging magnetic fields and ion-neutral collisions 
\cite{Judge2020mn}, so that \item all the elementary physical 
ingredients are naturally available
to generate a magnetically active corona  \cite{Parker1994}, then 
\item the pressure of hot coronal plasma leads to expansion that cannot be contained by the interstellar medium, causing the solar wind \cite{Parker1958}, and then 
\item the magnetic torque exerted by the Sun on the rotating, frozen-in wind plasma  then slows solar rotation, and finally  
\item on main sequence lifetimes, the slower  rotation weakens the  magnetic fields generated in the interior. 
\end{itemize}
%
 This story, based on first principle ideas.  is borne out
by data. Beginning with the work of Skumanich in 1975 \cite{Skumanich1972},
the rotation rates $\Omega$
of Sun-like stars were related to stellar age
$t$ as 
\begin{equation}
    \Omega(t) \propto t^{-1/2},
\end{equation}
based upon just four data points (open cluster stars and the Sun).  The recent advent of 
asteroseismological stellar age
determinations from
missions such as Kepler and TESS has vastly 
extended Skumanich's 
early picture, from 
a single relation $\Omega(t)$ to  dependencies on more variables.  As an example.  $\Omega(t,[B-V]_0)$, for main sequence stars, is shown in Figure~\ref{fig:gyro}.   

\begin{figure}[ht]
\begin{center}
\includegraphics[width=0.8\linewidth]{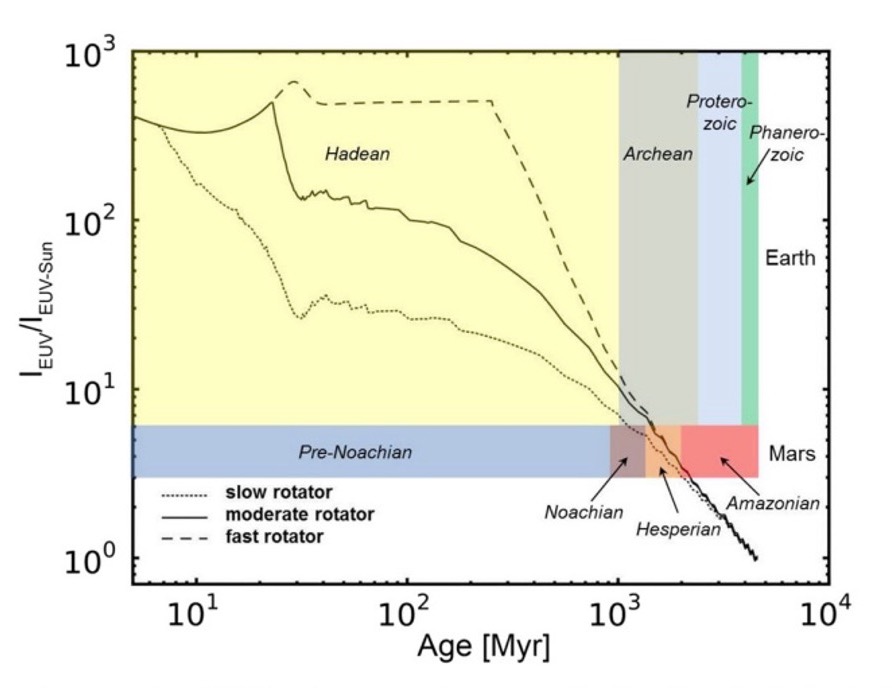}
\end{center}
\caption{ Three possible histories of high energy (EUV) radiation of the Sun are plotted, depending on the assumed zero age main sequence rotation rate.  The ``Skumanich'' law, derived 
from data between 40 and 4000
Myr, would be a straight line with slope -1/2 in this plot. Various eras in the histories of Earth and Mars are indicated. The earliest fossil evidence for life on Earth is 
dated at 3465 Myr ago, an age of $\approx 1080$ Myr, defining the beginning of the Archean era. 
Figure taken from \cite{Lammer+others2018}.
}
\label{fig:lammer}
\end{figure}

Armed with data from a variety of
spacecraft over decades, we
can consequently infer that the early Sun flooded the solar system with
far more intense UV and X- radiation than it does today.
Figure~\ref{fig:lammer} is 
a recent representation of 
possible histories of 
high energy radiation of Sun-like stars with age on
the main sequence. 
The implications of this 
story of solar rotation are
explored further in the next section.  

\section*{A closer look}

\subsection*{Stars like the Sun}
The Sun does not belong to any known   
stellar group, but it is similar to 
the stars in the open cluster M67.  
Gyro-chronology 
suggests an age of 4 Gyr for the stars of M67.  The Sun is 4.54 Gyr old (Table 1).
The cluster is too distant (800-900 pc) to observe G2~V stars 
at UV and  X-ray wavelengths  (m$_V\approx 14.5$).   Thus we are left to compare the Sun's magnetic activity mostly with nearer, brighter field stars. 
\begin{figure}[ht]
\includegraphics[width=0.9\linewidth]{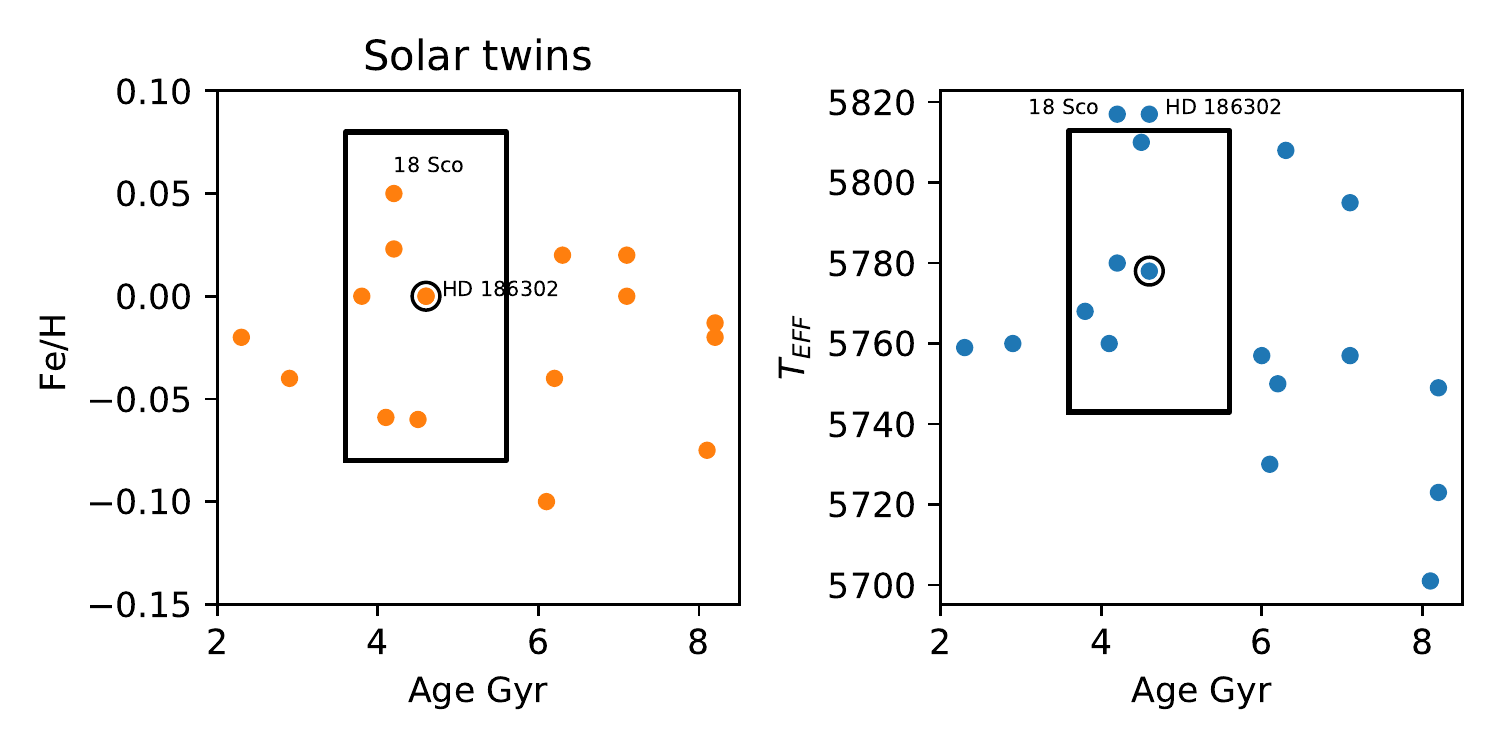}
\caption{Known solar twin candidates are plotted in terms of metallicity (left panel) and effective temperature (right panel) as a function of stellar age.  The Sun is shown as an open black circle. Estimates of typical uncertainties are represented by the black box. The data are taken from
\protect\url{https://en.wikipedia.org/wiki/Solar_analog}.}
\label{fig:twins}
\end{figure}

Many authors have sought a genuine ``twin'' for the Sun among the stars.  Currently the best candidates are HD 146233 (18 Sco) and   HD 186302
(https://en.wikipedia.org/wiki/Solar\_analog). 
A group of genuine ``solar twin'' would be
of enormous practical use and interest in our quest to relate solar behavior to other stars.  Genuine twins would permit us to assess if there are any special properties about the Sun.   However, nature presents us with challenges:
\begin{itemize}
    \item The Sun is ``old'' in the sense that spin-down by angular momentum loss has already occurred at 4.5 Gyr sufficient to have erased any ``memory'' of the ZAMS angular momentum. (Figure~\ref{fig:lammer}). As such it is therefore magnetically ``inactive'' among its younger stellar relatives, and it is not highly luminous in UV or X-ray wavelengths. 
    \item While there are  $\approx 500$ stars in our immediate  neighborhood ($d \leq 30$ pc), they reflect the known initial mass distributions and so of these, only about 25 are G stars of luminosity class V.  Of these only about 13 have 
    spectral types 
    between G0 and G4. 
    \item The dimensionality of a ``Sun-like'' space of variables is a little subjective.  But this space must contain at least ZAMS mass, metallicity, age and rotation rate. Figure~\ref{fig:twins} 
    reveals the dearth of possible candidates in two scatter plots.  
\end{itemize}
Not only are there few twin candidates, but the typical  uncertainties are large: 
 4-5 different stars are
 identical with the Sun 
 within the uncertainties.  But this situation is expected to improve as Sun-like stars 
become the focus of the large number of
exoplanetary scientists. 

\subsection*{Stars of solar mass through time}

\begin{figure}[ht]
\includegraphics[width=0.6\linewidth]{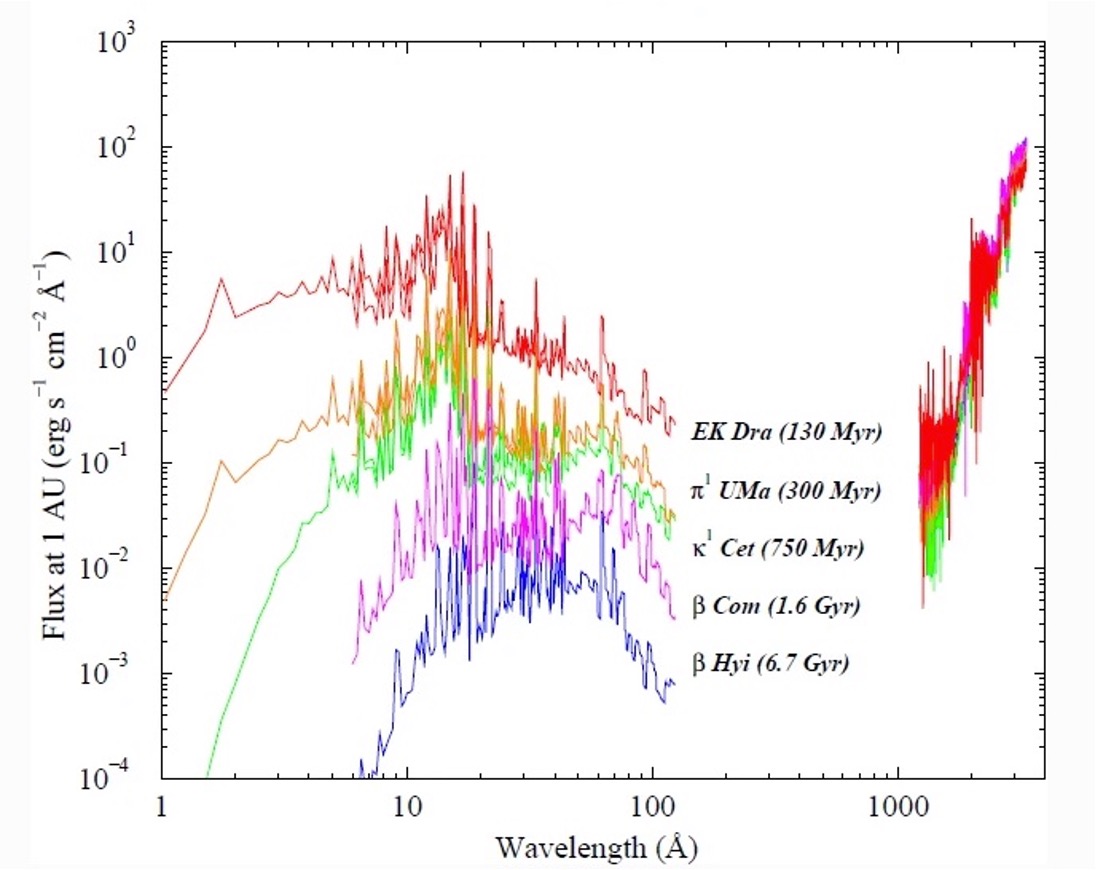}
\includegraphics[width=0.39\linewidth]{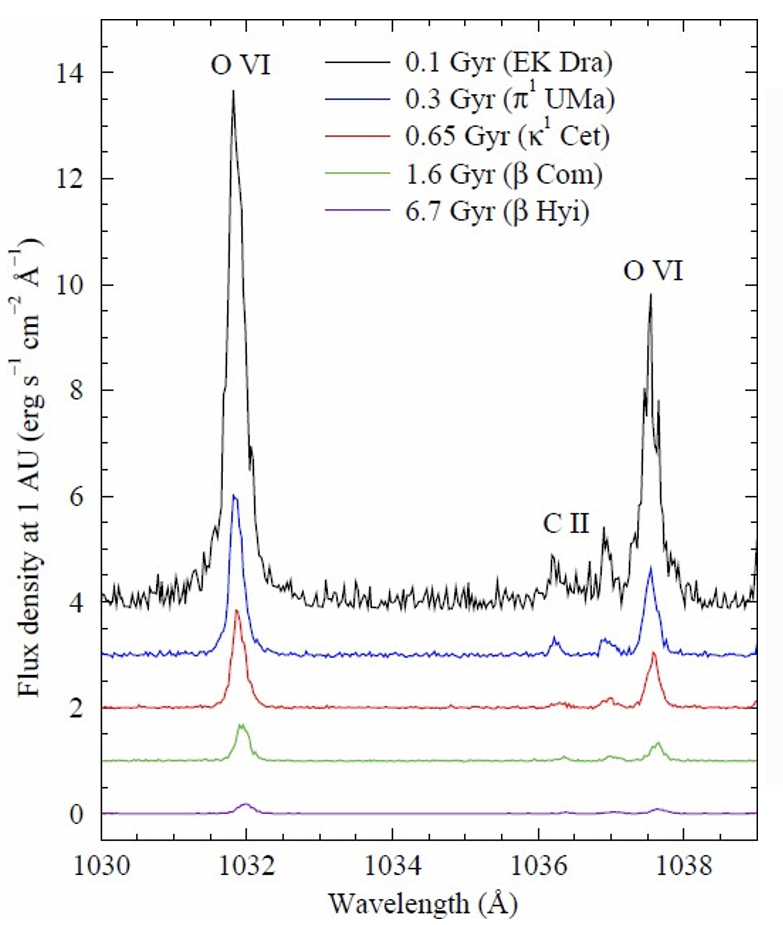}
\caption{Levels of  emission from Sun-like stars are shown 
as a function of age
on the main sequence.
Flux densities at the stellar surface can be
derived by multiplying 
by (1AU/$R_\odot$ = 215)$^2=4.6\cdot 10^{4}$.
Note the logarithmic and linear scales plotted, and the gap between 120 and 1000 \AA{} caused by 
interstellar absorption.  
Figures from \cite{Guedel2007}.
}
\label{fig:guedel}
\end{figure}

Again our focus is on magnetic activity and not
stellar evolution \textit{per se}.
Researchers have documented 
not only the high energy emission, but also bulk wind properties using observations at visible, UV to X- ray 
wavelengths. Nice reviews
are available from G\"udel \cite{Guedel2007,
Guedel2020SSRv..216..143G}.  Figure~\ref{fig:guedel} highlights how high and low energy emission evolves in time, as
represented by surface flux density measurements of several stars 
of 1$_\odot$.  The 
harder the radiation, 
the faster it decays
with age.
The generic ``EUV'' emission (wavelengths between about 10 and 91 nm) used in the geospace community cannot be measured for enough stars owing to large Lyman-continuum optical
depths in the interstellar medium. 
The ``EUV" emission presumably lies between
the UV and X-ray
behaviours shown.  
The very steep trends in UV emission and X-ray emission are
somewhat different, the latter indicating that coronal soft X-ray emission cannot exceed
a limit near $2\cdot10^{30}$ erg, $6\cdot10^7$ erg~cm$^{-2}$s$^{-1}$, or 
0.005 L$_\odot$.   This ``saturation'' has been known for almost 4 decades, but 
the underlying 
reasons are still debated. 
Note that the present day Sun's 100nm UV spectrum (emission lines and 
continua are both important for
upper atmospheric chemistry and
dynamics) is some 1 and 2 orders of
magnitude weaker than 
stars of age 2 Gyr and 0.1 Gyr 
respectively.  

The
(non-linear) dynamical responses of 
the Earth's atmosphere to
between 1 and 10 times the 
mean solar EUV flux 
have been studied
\cite{Tian+others2008,Volkov2016MNRAS}. The Earth's 
atmosphere does not blow off
(i.e. the thermal energy at the ``exobase'' does
not exceed a critical fraction of the gravitational potential energy). This is because hydrodynamic flow and adiabatic expansion 
sap the available energy for heating the upper atmosphere for levels of EUV flux 5 times those of the mean present-day Sun. 
In contrast Mars would have lost any initial atmosphere, only able to
maintain a warm and wet period several hundred Myr after Mars formed (see Figure~\ref{fig:lammer}) when the EUV fluxes dropped 
within an order of magnitude of
the current Sun \cite{Tian+others2009}.

\subsection*{Flares and CMEs}

\begin{figure}[ht]
\begin{center}\includegraphics[width=0.7\linewidth]{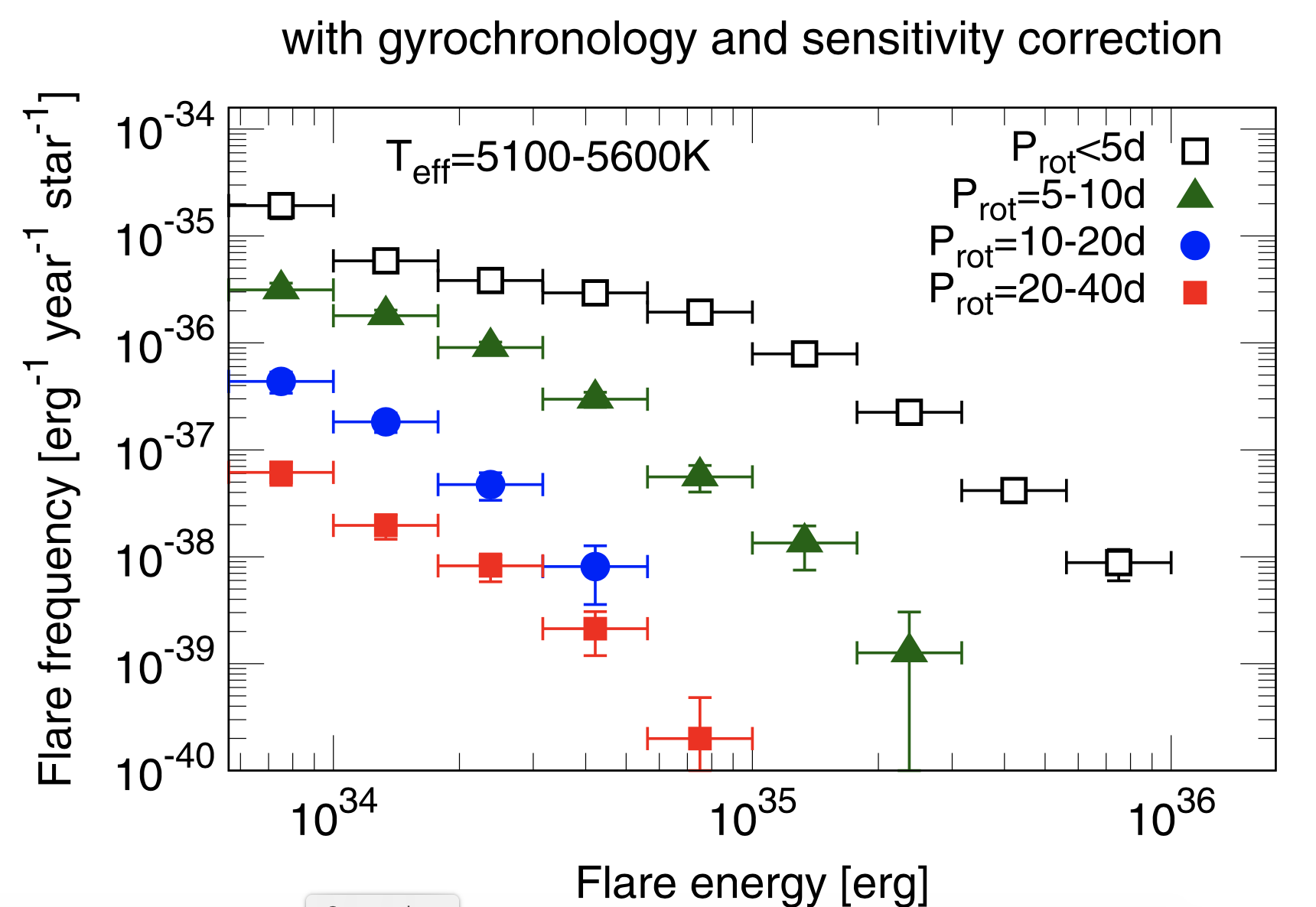}
\end{center}
\caption{The occurences of flares on Sun-like stars from the Kepler mission are shown as a function of flare energy in several stellar age bins.  The strongest ever (``Carrington'') flare of 1859 
is estimated to have released about $10^{33}$ erg.  The figure suggests that
a flare of $10^{34}$ erg might occur once
every 2000 years or so
on the present Sun.   
In contrast, at 1 Gyr
(a rotation period 
near 10 days) this flare would
occur once every 30 years or so.
Figure from \cite{Okamoto+others2021}.}
\label{fig:flares}
\end{figure}

The above arguments rely on data from  
``typical'' observations, not those rarer phenomena such as flares. Again, a vast quantity of data has been analyzed for the Sun \cite{Fletcher+others2011} and
significant progress again has been
made with recent 
photometric asteroseismology missions for sun-like stars \cite{Okamoto+others2021}. One might expect that flaring might higher in 
intensity and frequency 
in younger stars.  The story is even now unfolding as recent satellite databases undergo more and more scrutiny.   Flares recorded on
Sun-like stars (Figure~\ref{fig:flares}) extend
far higher in energy than the
largest in 
solar history \cite{Cliver+Dietrich2013}, the ``Carrington Event'' of 1959, with an estimated energy of $10^{33}$ ergs.  Flares last shorter than 24 hours, so that while the impact of flare radiation on surface life might be serious 
on the illuminated hemisphere, areas in shadow would not
receive devastating 
doses of high energy radiation.   We can speculate that this intermittent source of energy on 
genetic mutations might have been relevant to
the evolution of complex life on Earth.

\section*{Conclusions}

Absent from this brief  discussion is the importance of
understanding why the Sun is obliged to produce sunspots with the pulse of a 
22 year 
cycle
(Figures~\ref{fig:cycle}, \ref{fig:hathaway}).  Hopefully ground-based observations of 
chromospheric Ca$^+$ lines 
begun in 1966 will continue over many more decades \cite{Egeland2017} and for more stars, to improve our knowledge of what might cause and 
then suppress cycling behavior, for stars having the same 
convection and rotation properties.  The answer to why the Sun and a few stars must do this is central to our understanding of magnetic evolution, also giving insight into the weaker pulse of  of sunspot signals during the Maunder Minumum (1645-1715, 
\cite{Eddy1976}).
Curiously, the 
Sun presents the most
ordered 22-year cycle of all field G stars, only about
10\%{} of which 
show clear cycles
\cite{Egeland2017}.
Young field G stars tend to show strong irregular variability, old 
stars weak, if any, secular variations. 
In terms of cycle properties, the Sun is more similar to field K stars.

The magnetic machinery of Sun-like stars 
is only partly understood from first principles.  Neither the operation modes of the global dynamo or the 
manner in which energy is released in the atmosphere with accompanying high energy particles and radiation are understood.  They are 
relatively well documented in our observations.  But because of the enormous challenges  
facing theorists, 
solar physics will continue to be 
an observationally-driven field.  The place of the Sun among the stars will remain of central interest to our understanding of astrophysical dynamos and plasma physics, and it will 
become clearer as our datasets improve.

\bibliographystyle{ws-fnl}
\bibliography{biblio}

\input theory.tex

\end{document}

%% file: theory.tex
\section*{Appendix: The Sun's  magnetic engine}

There is no reason \textit{a priori}  that the Sun should behave in this fashion.   Although
the required power of all this ``activity'' is a small fraction  of
the total solar luminosity, the oscillatory behaviour has garnered attention 
of physicists.  This behavior is striking  because the implied magnetic \textit{order} exists 
in spite of the fact that energy transport across the outer 28\%{} of the Sun, by radius, is dominated by
turbulent convection. Perhaps solar magnetism is a manifestation of
\textit{emergent} behavior arising from the complexity of a complex dynamical system \cite{Holland2014}.
Alternatively, a more deterministic 
mechanism might be in action, 
perturbed by convection to produce
the ``noise'' seen, for example, in  Figure~\ref{fig:hathaway}, and others.  The latter picture is almost universally adopted 
by solar physicists, and is adopted below because it has value
pedagogically.   

%
\begin{figure}[ht] 
\begin{center}
\includegraphics[width=0.6\linewidth]{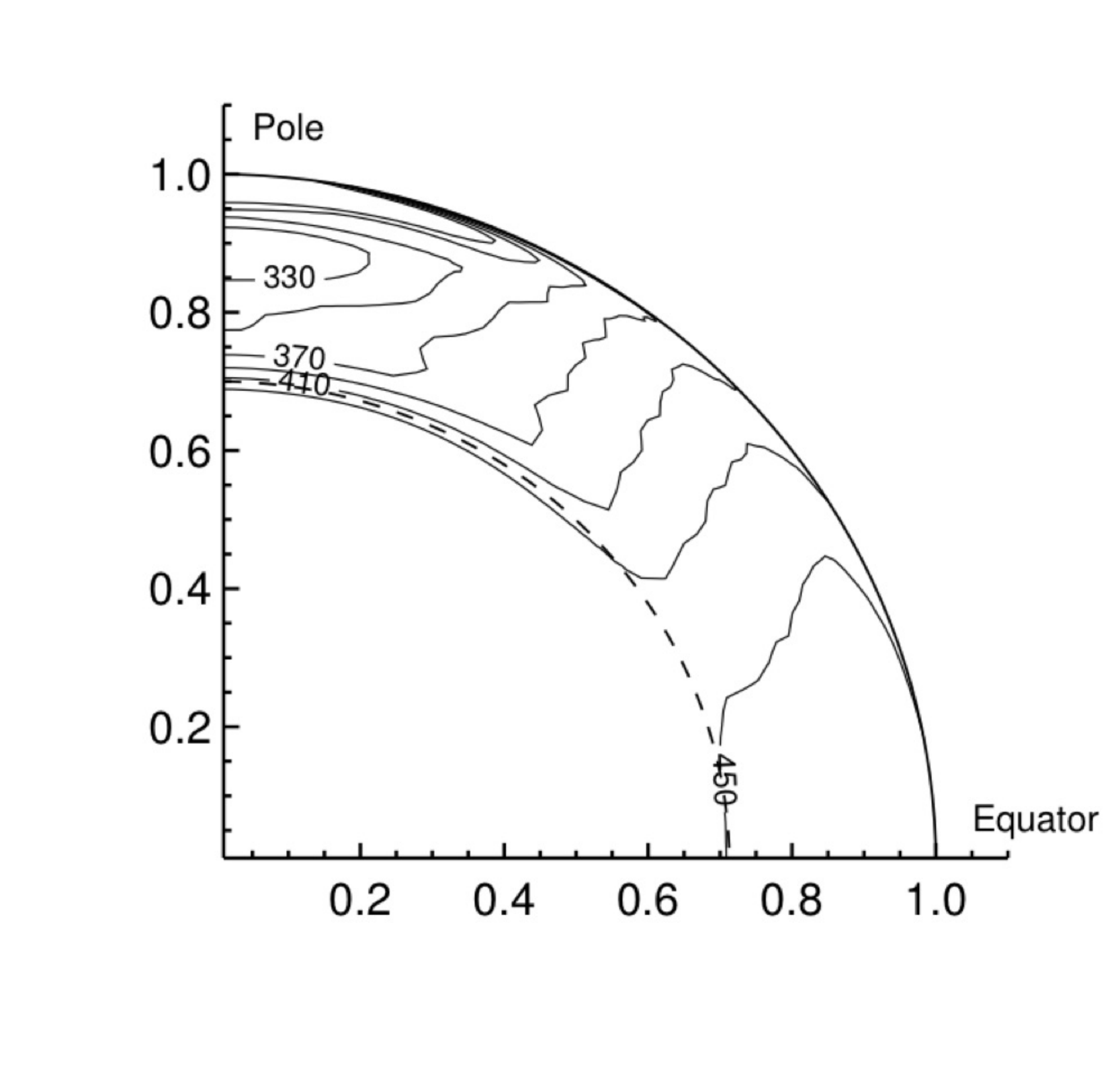}
\end{center}
\caption[Differential rotation]{The figure shows the internal rotation
  rates \textit{in the toroidal direction} of the solar interior as a
  function of radius and latitude as derived from helioseismology.
  The lines are contours of constant rotation rate in units of
  10$^{-9}$ Hz.  The closely packed contours in
  the figure are favourable locations for the amplification of magnetic fields
  through the $\Omega$-effect.  Data 
  from \cite{Judge2020}.} \protect\label{fig:helios}
\end{figure}
Much  observational evidence 
and physical considerations  strongly
argue in favor of 
the regeneration of magnetic fields within the solar interior
(e.g., \cite{Brun+Browning2017}.)
Since 1989, one critical ingredient
of deterministic models 
has been measured through 
observations of 
the modulation of the Sun's normal modes of oscillation  (``helioseismology").  
Until measurement of 
the internal profile for  axial differential rotation ${\Omega}_\varphi(r,\vartheta)$ as a function of radius $r$ and
latitude $\vartheta$ were available, only the \textit{surface} and coronal rotation properties were
accessible  to observers. Helioseismology has identified three interior regions The solar interior has three shear layers  in the interior
(Figure~\ref{fig:helios}).  
Radial shears in $\Omega_\varphi$ are found just below 
the convection zone (the ``tachocline'') and just beneath the photosphere.  A latitudinal 
shear is found near latitudes of $|\vartheta| = 55^\circ$.   These large-scale
shear zones can readily generate 
toroidal magnetic fields by stretching magnetic field lines around the axis of rotation. Called the $\Omega$-effect,  these shears are believed to be 
a critical component of credible
large-scale,  dynamos 
(see Figure~\ref{fig:cycle}).   
But in order to make the Sun's magnetism oscillate, 
another effect is needed, and one
example of a model is discussed 
briefly here, the $\alpha$-effect.

To complete a deterministic model requires ingredients in addition to the helioseismic large-scale shear motions (e.g. \cite{Parker1955}):
\begin{itemize}
\item  sources and  sinks of magnetic fields are needed to produce repetition over 22 years,
\item processes that convert 
poloidal fields to 
toroidal fields are needed, which must
    \item break cylindrical symmetry  \cite{COWLING1933}.     \end{itemize}
Asymmetries in the internal fluid dynamics arise directly from  rotation, because the Coriolis force ($-2\rho~\bm{\Omega}\times \vec{u}$) that acts upon convective flows of density $\rho$ and velocity $\vec{u}$, 
is a pseudo-vector, i.e. asymmetric under reflection.   
To proceed further, we can consider the Sun as a magneto-hydrodynamic (MHD) system
(e.g. \cite{Parker2009}). The magnetic field then evolves according to the \textit{induction equation}\footnote{A combination of Ohm's law, Faraday's law of electromagnetic induction, Amp\'ere's law and 
the lowest order transformation 
of electric fields in the frame of
the fluid moving with velocity \textbf{u})}:
\begin{equation} \label{jeqinduction}
{ {\partial \vec{B}}\over {\partial t}} = \curl (\vec{u} \times \vec{B}) +
\eta \nabla^2 \vec{B}, \ \ \eta = 1/\mu_0\cond,
\end{equation}
where $\cond$ is a scalar electrical conductivity and $\mu_0$ the permeability of free space.   
To zeroth order the  observations indicate that global surface magnetic fields (the left hand side of 
equation~\ref{jeqinduction})
must change on time scales 
of a decade.   
Kinetic theory gives us fluid transport coefficients such as conductivities 
$\sigma$ \cite{Braginskii1965}, from which 
$\eta \sim 1$ m$^{2}$~s$^{-1}$ in the Sun's interior.  Choosing
$\ell \sim R_\odot/3=2\times10^{8}$ cm,  the diffusive term in
equation~(\pref{jeqinduction}) has a time scale of
$\ell^2/\eta \sim 10^{9}$ years!  Evidently, the ``advective'' term
$\curl (\vec{u} \times \vec{B})$
must not only provide a source, but also must drive a sink for magnetic fields when integrated over the volume of the Sun. 
Correlations and anti-correlations between \vec{u} and \vec{B} can lead both to positive and negative contributions to equation~(\ref{jeqinduction}).  Such correlations must operate on
 spatial scales $\ll R_\odot$ in order to
produce time scales of years, but must have consequences on global length scales.   Inspired by Parker's 
notion of small-scale convective cyclonic turbulence  \cite{Parker1955} 
one class of dynamo model seeks solutions to the development of large-scale vector fields $\overline{\vec{X}}$, $\vec{X}= 
\overline{\vec{X}} + \pr{\vec{X}}$, where the vector field $\vec{X}$ is assumed to be consist of  separate 
large and small scales $\overline{\vec{X}}$ and  $\pr{\vec{X}}$.
The  small-scale  correlations are averaged out and
written in terms of tensor coefficients, and the reader is referred to  an excellent review by Rempel  
summarizing further ideas
\cite{Rempel2009}.   Of the various
tensor coefficients, the most important here is 
$\alpha$ which appearing as a source term
in the poloidal magnetic field
which is otherwise absent
(e.g., \cite{1999ApJ...518..508D}). 
\begin{equation}
  \frac{\dd\avr{\B}}{\dt}=\ldots + \curl\left(\alpha\avr{\B}\right)=
  \ldots + \alpha\mu_0\avr{\jj}\;,
\end{equation}
i.e. the induced magnetic field induced by motions contributing to
$\alpha$ is proportional to the mean current, completing the ``circuit''
shown in Figure~\ref{fig:cycle}. 
Rempel 
\cite{Rempel2009} shows that, 
when stratification exists (under a gravity vector $\vec{g}$), then 
\begin{equation}
  \alpha\approx\alpha_0(\vec{g}\cdot{\boldsymbol \Omega})=\tau_c^2v_{\rm rms}^2
  {\boldsymbol \Omega}\cdot\grad\ln(\varrho v_{\rm rms}),
\end{equation} 
where the rms turbulent speed $u_{rms}$ and turnover time $\tau_c$
characterize the turbulence, and $\alpha_0$ is a constant.   
Note that $Ro = (\Omega\tau_c)^{-1}$ is the well-known  ``Rossby'' number.  In short, 
we have physical ingredients to generate a cycling dynamo,
summarized as follows 
(see Figure~\ref{fig:cycle}):
\begin{equation}
  \B_p\stackrel{\Omega}{\longrightarrow}\B_t 
  \stackrel{\alpha}{\longrightarrow}\B_p\stackrel{\Omega}{\longrightarrow}\B_t\ldots,
  \label{alpha2-cyc}
\end{equation}
where both the $\alpha$ and $\Omega$ model parameters depend on the timescales for rotation, differential rotation and 
convection. 
In summary, 
observations can constrain physical models of evolving solar
magnetism, suggesting that the quasi-deterministic picture 
has some merit.  The native asymmetries, the roles of rotation and differential rotation 
are all essential components that 
are reasonably well understood. However helioseismology, while resolving 
the solar interior, 
has also shown more recently that subsurface convective motions $u_{rms}$ are 
at least an order of magnitude weaker (at $r\approx 0.96R_\odot$) than theory 
suggests \cite{2012PNAS..10911928H,2021A&A...649A..59B}, which if confirmed begs the worrying question of what transports the solar luminosity there. Nevertheless it must generate some $\alpha$-effect and link large- and small-
scales through non-linear terms 
that may be deterministic only in 
a global (i.e. highly-averaged) sense.  For more details 
including the fascinating global problem of secular hemispheric 
accumulation of magnetic 
helicity in such models \cite{Low1994}, and 
non- mean-field models, such as the 
Babcock-Leighton picture where the $\alpha$ effect 
takes place primarily in surface dynamics,
the reader can 
should consult
modern reviews, for example  \cite{Rempel2009,Charbonneau2020}.